\documentclass[12pt, preprint]{aastex}

\usepackage{graphicx}

\newcommand{\sbu}{erg s$^{-1}$ cm$^{-2}$ arcsec$^{-2}$}
\newcommand{\flu}{erg s$^{-1}$ cm$^{-2}$ arcsec$^{-1}$}

\newcommand{\kms}{km s$^{-1}$}
\newcommand{\mum}{$\mu$m}
\newcommand{\cc}{cm$^{-3}$}

\newcommand{\spitzer}{\textit{Spitzer}}
\newcommand{\chandra}{\textit{Chandra}}

\newcommand{\nii}{[\ion{N}{2}]}
\newcommand{\oiii}{[\ion{O}{3}]}

\slugcomment{Accepted for publication in the Astrophysical Journal}

\shorttitle{Kepler SNR 2nd epoch HST observations}
\shortauthors{Sankrit et al.}

\begin{document}

\title{Second Epoch Hubble Space Telescope Observations of Kepler's
Supernova Remnant: The Proper Motions of Balmer Filaments\footnotemark[1]}

\footnotetext[1]{Based on observations made with the Hubble Space Telescope.}

\author{Ravi Sankrit\altaffilmark{2},
John C. Raymond\altaffilmark{3},
William P. Blair\altaffilmark{4},
Knox S. Long\altaffilmark{5},
Brian J. Williams\altaffilmark{6},
Kazimierz J. Borkowski\altaffilmark{7},
Daniel J. Patnaude\altaffilmark{3},
\and
Stephen P. Reynolds\altaffilmark{7}
}

\altaffiltext{2}{SOFIA Science Center, NASA Ames Research Center, M/S N211-3, Moffett Field, CA 94035.}
\altaffiltext{3}{Smithsonian Astrophysical Observatory}
\altaffiltext{4}{Johns Hopkins University}
\altaffiltext{5}{Space Telescope Science Institute}
\altaffiltext{6}{CRESST/USRA and X-ray Astrophysics Laboratory, NASA GSFC}
\altaffiltext{7}{North Carolina State University}

\begin{abstract}

We report on the proper motions of Balmer-dominated filaments in
Kepler's supernova remnant using high resolution images obtained
with the Hubble Space Telescope at two epochs separated by about
10 years.  We use the improved proper motion measurements and revised
values of shock velocities to derive a distance to Kepler of
$5.1^{+0.8}_{-0.7}$~kpc.  The main shock around the northern rim
of the remnant has a typical speed of 1690~\kms\ and is encountering
material with densities of about 8\,\cc.  We find evidence for the
variation of shock properties over small spatial scales, including
differences in the driving pressures as the shock wraps around a
curved cloud surface.  We find that the Balmer filaments ahead of
the ejecta knot on the northwest boundary of the remnant are becoming
fainter and more diffuse.  We also find that the Balmer filaments
associated with circumstellar material in the interior regions of
the remnant are due to shocks with significantly lower velocities
and that the brightness variations among these filaments trace the
density distribution of the material, which may have a disk-like
geometry.

\end{abstract}

\keywords{ISM:individual objects(Kepler, SN1604, G4.5+6.8) -- ISM:supernova remnants}


\section{Introduction}

The historical supernova SN1604 was observed and documented by
Johannes Kepler \citep{field77}.  The supernova remnant (SNR)
resulting from the explosion was discovered in a targeted search
using the 100-inch reflector on Mt.\ Wilson ``in the expectation
that the ejected masses would still be visible" \citep{baade43}.

Kepler's SNR (G4.5+6.8, hereafter Kepler) is located well away from
the Galactic Plane, at a height of $590d_5$ pc, where $d_5$ is the
distance to the remnant in units of 5~kpc.  It shows up as a
well-defined shell with a radius of about $2.6d_5$ pc in radio,
X-ray \citep{delaney02} and infrared 24\mum\ band \citep{blair07}
images.  There is a pronounced asymmetry in the emission at these
wavelengths, with the northern limb being much brighter than the
southern limb.  The contrast is more extreme in the optical, where
filaments and knots extend across the northern and northwestern
limb while the southern limb is not seen at all \citep{blair91}.
The brightest optical emission is from a region in the northwest.
Spectra of this bright emission show radiative shock emission arising
from material with densities $n_{e}$ of $\sim$1000\,\cc\
\citep{dennefeld82, leibowitz83}.  These densities are much higher
than expected for the interstellar medium at the location of Kepler,
well off the plane of the galaxy, and therefore these measurements
have been interpreted as evidence that the emitting gas is from the
circumstellar medium (CSM) of the supernova progenitor system
\citep{blair91}.

The supernova was classified as a Type I by \citet{baade43} based
on the reconstructed historical light curve, but that conclusion
was called into question by \citet{doggett85}, who found that the
light curve was equally consistent with a Type II-L supernova.  The
presence of nitrogen-rich circumstellar material (enhanced by
$\sim$0.5 dex) appeared to favor a core-collapse origin \citep[e.g.][]
{bandiera87, borkowski92}. However, \citet{blair07} pointed out
that much of this nitrogen enhancement could arise simply from the
galactic abundance gradient in the direction of Kepler, almost
directly toward the galactic center. Its distance from the Galactic
plane and X-ray spectral results \citep[e.g.][] {kinugasa99,
cassamchenai04} suggested that a Type Ia supernova was more likely.
The debate over the supernova type was settled only with the analysis
of a long (750~ksec) \textit{Chandra} observation by \citet{reynolds07}.
Based primarily on the strong Fe emission and lack of O emission
from the ejecta, they confirmed that Kepler was the result of a
Type Ia supernova.  More recently, \citet{yamaguchi14} reconfirmed
the Type Ia origin for Kepler based on the Fe K-shell emission
observed with \textit{Suzaku}.

In a few locations in Kepler, coincident with the bright radiative
emission, \citet{reynolds07} detected oxygen emission and found
solar O/Fe abundance ratios, which they associated with the shocked
CSM\@.  \citet{burkey13} carried out a detailed analysis of the
\textit{Chandra} data and found that the circumstellar material is
primarily distributed around the bright north rim, but that substantial
amounts of it are found projected near the center of the remnant.
They follow \citet{bandiera87} in explaining the bright northern
rim as due to material piled up by the northward motion of the
progenitor system, and they explain the centrally-projected CSM as
a disk or torus of material seen edge-on.

The confirmation of its Type Ia origin, and the presence of the CSM
imply that the progenitor system underwent significant mass loss
prior to the explosion, and that the explosion most likely followed
the single-degenerate scenario in which a white dwarf accretes
matter from a non-degenerate companion, rather than the double-degenerate
scenario in which two white dwarfs merge.  Based on the structure
and properties of the shocked CSM, several groups \citep{chiotellis12,
burkey13, toledoroy14} have modeled the progenitor system as a
symbiotic binary, with the white dwarf accreting from an AGB star.
In this scenario, the AGB wind shapes the CSM around the supernova.
Strong silicate features detected in \spitzer\ IRS spectra of Kepler
provide observational support for an AGB progenitor-companion as
well \citep{williams12}.  This companion, which should have survived
the explosion, has so far not been discovered \citep{kerzendorf14}

The distance to Kepler continues to be a matter of debate.  Using
a kinematic model of the galaxy, \citet{reynoso99} placed a lower
limit of $4.8\pm1.4$~kpc based on an H~I absorption feature, and
an independent upper limit of 6.4~kpc based on the lack of H~I
absorption from a known cloud.  Our earlier work, described further
below, used the proper motion of a Balmer filament and an estimated
shock velocity to derive $d = 3.9^{+1.4}_{-0.9}$~kpc \citep{sankrit05}.
In recent years, distances of either 4 or 5~kpc have been assumed
in almost all the published literature on Kepler.  However, simulations
that combine the evolution of ejecta and the CSM seem to require
that the distance be greater than 6~kpc if the explosion energy were
$10^{51}$~ergs, the standard value for a Type Ia supernova
\citep{chiotellis12}; a smaller distance to the remnant
would require a sub-energetic Type Ia explosion.  \citet{patnaude12}
find that the amount of Fe in the X-ray spectrum indicates a larger
than normal explosion energy for Kepler, which coupled with the
CSM density profile in their model requires that the remnant be
at a distance of at least 7~kpc.

Optical hydrogen Balmer filaments arise in regions just behind fast
so-called non-radiative shock fronts where the pre-shock material
is partly neutral.  Charge exchange between neutral hydrogen atoms,
which are unaffected by the shock, and shock-heated protons results
in a population of fast neutrals.  Some fraction of these atoms are
excited and emit line photons before getting ionized \citep{chevalier78}.
At this stage, the post-shock gas has not had sufficient time to
recombine and cool via the emission of collisionally excited forbidden
lines, and therefore, the strongest optical line by far is Balmer
H$\alpha$.  The H$\alpha$ line thus has narrow and broad components
arising from the neutrals at the pre-shock and post-shock gas
temperatures, respectively.  The width of the broad component is
related to the shock velocity, albeit with various complications
that have been examined recently \citep{morlino13, shimoda15}.

Spectra of Balmer filaments in Kepler were obtained by \citet{fesen89}
and \citet{blair91}, providing shock velocity estimates for a few
positions. Proper motions of Balmer filaments have been measured
by \citet{sankrit05}, based on a comparison of a ground based image
obtained in 1987 with one from the Hubble Space Telescope (HST)
obtained in 2003.  The low angular resolution of the ground based
image and the crowded star field towards Kepler permitted us to
measure the motion only at two locations near each other along a
single filament in the northwest.  In this paper, we present improved
proper motion measurements of many Balmer filaments all around the
SNR, using a second epoch of high angular resolution HST data with
the first epoch HST data obtained nearly 10 years earlier.  These
measurements are used to improve the distance estimate to Kepler's
SNR.

The second epoch observations and the procedure for aligning both
epochs of images are described in \S2.  A brief overview of the
optical emission from Kepler is given in \S3, and the proper motion
measurements of the Balmer filaments are presented in \S4.  The
results and their implications for our understanding of Kepler's
SNR are discussed in \S5, and finally \S6 contains a summary of the
main results.


\section{Observations, Data Processing and Image Alignment}

Kepler was observed with the HST Wide Field Camera 3 (WFC3) between
July 1 and 3, 2013, as part of a Cycle 20 Guest Observer Program
(PID 12885).  Observations were obtained in the UVIS channel through
the narrow band filters F656N (9870~s) isolating H$\alpha$ emission
and F658N (3894~s) isolating \nii\ $\lambda$6584 emission.  Exposures
through the largely line-free F547M filter were obtained and used
for star alignment and astrometry. The images were centered on
$\alpha_{J2000}, \delta_{J2000}$ = 17:30:40.8, $-$21:28:53.40.  The
WFC3/UVIS field of view, 162\arcsec\ $\times$162\arcsec\ is sufficient
to include most of the optical emission from the remnant; only the
eastern extremity of the northern limb, and the very faint filaments
on the extreme western edge are outside the observed field of view.
The observations were obtained using a standard dither and offset
pattern in order to remove cosmic rays and hot pixels, fill in the
chip gaps, and to completely sample the PSFs.  The Charge Transfer
Efficiency (CTE) losses were minimized using an appropriate post-flash
for each exposure to reach a pedestal of 12 electrons.

The data were processed using the DrizzlePac software suite
\citep[e.g.,][]{fruchter10,gonzaga12}, as provided by STScI\footnote{See
http://drizzlepac.stsci.edu.}.  The exposures through each of the
filters were corrected for CTE losses and then combined using the
AstroDrizzle package.  Alignment of the images was done using the
\textsc{tweakreg} routine along with custom SExtractor star catalogs
in each case.  For images with a given filter, the RMS residuals
are typically less than 0.05 pixels.  In order to provide an absolute
world coordinate system (WCS) frame, the WFC3 F547M was chosen as
the reference filter and the drizzled image aligned to the 2MASS
point source catalog \footnote{available at
http://www.ipac.caltech.edu/2mass/releases/allsky/}.  The best fit
gave RMS residuals of about 0.5 pixels (0.02\arcsec\,).  The remaining
WFC3 drizzled images were aligned with the F547M image, and the RMS
residuals were about 0.05 pixels.

The first epoch HST observations were obtained with the Advanced
Camera for Surveys (ACS) Wide Field Camera and have been described
by \citet{sankrit08}.  Briefly, they consist of observations obtained
through the following filters - F502N (\oiii\ $\lambda$5007), F658N
(H$\alpha$+\nii\ $\lambda$6584), F660N (\nii\ $\lambda$6584) and
F550M (line-free continuum).  The ACS $202\arcsec \times\, 202\arcsec$
field of view is larger than the WFC3 field, and was oriented
differently. Also, the ACS/WFC pixels are 0.05\arcsec.  The initial
step to bring the images from both epochs to a common frame was to
align the ACS F550M image with the WFC3 F547M reference image.
Then, the remaining ACS images were aligned to the F550M image.
Finally, the WCS solutions from the alignment procedures were
propagated back using the \textsc{tweakback} routine and all images
(epoch 1 and 2) were drizzled to a common $8500\times 8500$ pix$^{2}$
grid with a WFC3/UVIS native pixel scale of 0.0396\arcsec, rectified
to north up and east to the left.  The images resulting from this
final step are used in the figures, measurements and analyses
presented in this paper.


\section{Overview of the Optical Emission}

\begin{figure}
\begin{center}
\includegraphics[width=7.5in]{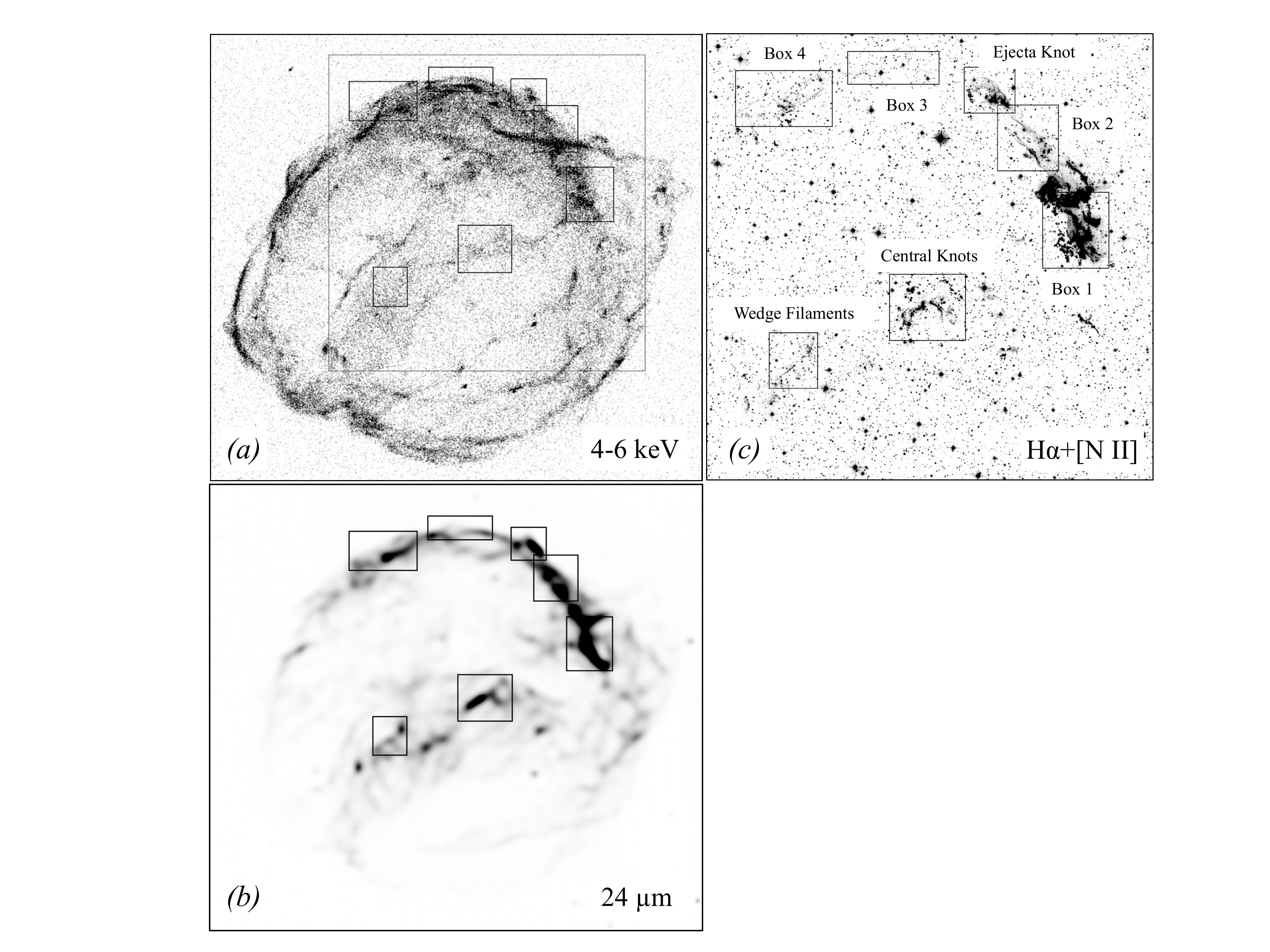}
\caption{\textit{(a):} \chandra\ image of Kepler.  The 4--6 keV
channel shows primarily the non-thermal emission.  The field of
view is 270\arcsec $\times$ 246\arcsec, with north up and east
to the left.
\textit{(b):} Deconvolved \spitzer\ MIPS 24$\mu$m image, with
the same field of view as the X-ray image.
\textit{(c):} HST/ACS F658N (H$\alpha$+\nii\,) image of Kepler 
obtained in 2003.  The field of view is 175\arcsec $\times$175\arcsec\
and corresponds to the larger box shown on the X-ray image.
In all three panels, the smaller boxes outline regions of interest
that will be discussed in this paper.  We will refer to these
regions using the labels shown on the optical image.
\label{fxacsirfull} }
\end{center}
\end{figure}

The optical emission from Kepler is sparsely distributed.  This is
clearly seen by comparing X-ray and optical images of the remnant.
Fig.~\ref{fxacsirfull}a shows a \chandra\ image in the 4--6\,keV
band.  The $270\arcsec \times\, 246\arcsec$ field includes the
entire remnant, and a near-complete shell is seen around its
periphery.  The emission in the band is predominantly non-thermal
and arises from the shock interaction between the supernova blast
wave and the interstellar medium (ISM).  Fig.~\ref{fxacsirfull}b
is a \spitzer\ MIPS 24\,$\mu$m image, deconvolved to an angular
resolution $\approx$ 2\arcsec\ using \textsc{icore} \citep{masci13},
and scaled to show the brightest emission.  The larger box on the
X-ray image outlines a trimmed $175\arcsec \times\,175\arcsec$ field
for which the ACS F658N image is shown in Fig.~\ref{fxacsirfull}c.
In all three panels, the smaller boxes outline regions that will
be discussed in the following sections.

The ACS image includes virtually all of the previously known optically
emitting filaments and knots in Kepler (a few lie outside the field
of view to the east-northeast and the west).  The optical features
consist of bright radiatively shocked knots that emit strongly in
both H$\alpha$ and \nii\ and fainter non-radiative filaments that
emit H$\alpha$ only.  The radiative emission is strongest in the
region labeled ``Box 1'', where the blast wave is running into
dense, clumpy circumstellar material.  The fainter Balmer filaments
produced by non-radiative shocks running into the ISM are seen along
the northern rim.  Spectra of the Balmer filaments \citep{fesen89,
blair91} show H$\alpha$ line profiles consisting of narrow and broad
components (FWHM $\approx$ 1750~\kms\,) confirming that the emission
is due to shock excitation.  Also along the rim is the shock created
by an ejecta knot that has punched through the primary shock.  The
regions designated as ``Wedge Filaments'' and ``Central Knots''
contain emission from both radiative and non-radiative shocks.
These regions appear to be part of a band running across the middle
of the remnant, NW to SE\@.  The band is clearly visible in the
broad-band X-ray image presented by \citet{burkey13}, who identified
much of the emission with CSM material, and can be discerned in the
\chandra\ image shown here in Fig.~\ref{fxacsirfull}.

In the following sections we will focus on the evolution of the
optically emitting material based on the two epochs of HST data.
A thorough discussion of the optical emission from Kepler, including
references to past work may be found in \citet{sankrit08}.


\section{Proper Motion Measurements}

\begin{figure}
\begin{center}
\includegraphics[width=6.5in]{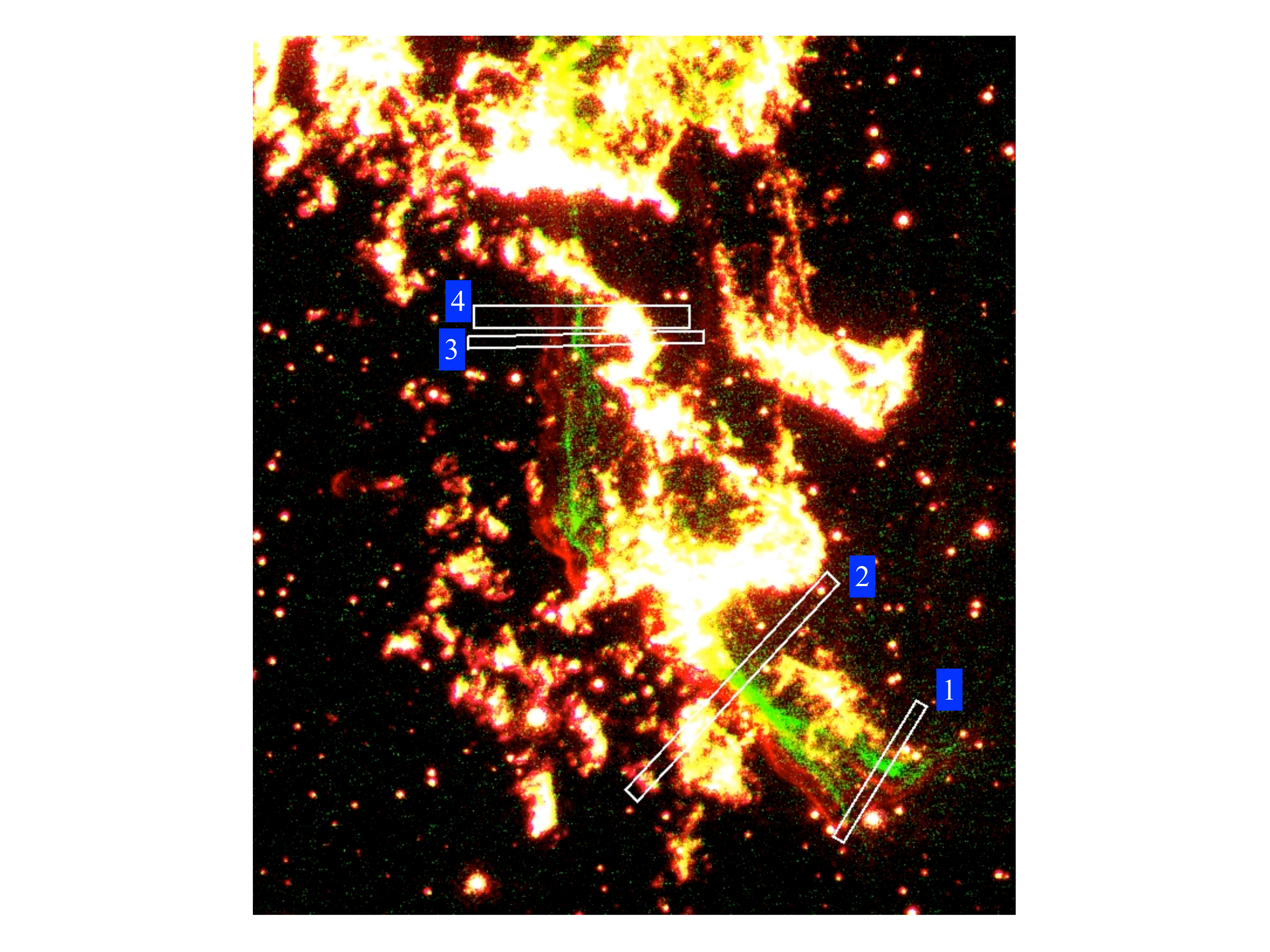}
\caption{Three-color image of the Box 1 region.  The ACS F658N image
showing epoch 1 H$\alpha$+\nii\ is in red, the WFC3 F656N image
showing epoch 2 H$\alpha$ is in green, and the ACS F660N image
showing epoch 1 \nii\ only is in blue.  The red and green features
are the Balmer filaments (only H$\alpha$ emission) in epoch 1 and
epoch 2, respectively.  Radiative shocks show up as white, and in
some regions as yellow due to the stretch used in the display.
Stars are white.  The cuts along which the proper motions were
measured are shown along with labels used in the text and in Table
1.  The image dimensions are about $26\arcsec\times 30\arcsec$.
\label{fbox1}}
\end{center}
\end{figure}

\begin{figure}
\begin{center}
\includegraphics[width=6.5in]{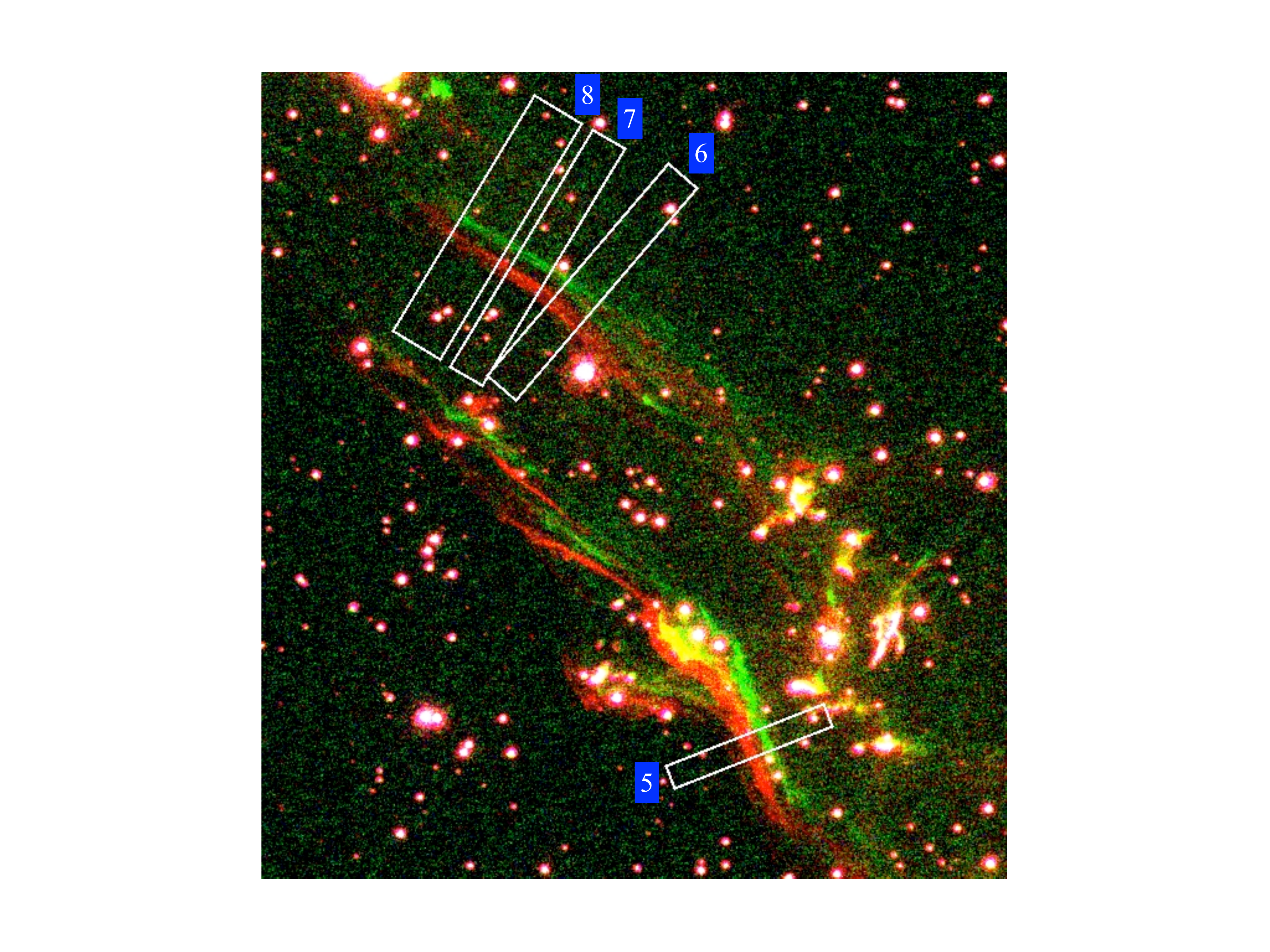}
\caption{The same as Figure \protect\ref{fbox1} for Box 2.  The
image dimensions are about $24\arcsec\times 26\arcsec$.
\label{fbox2}}
\end{center}
\end{figure}

\begin{figure}
\begin{center}
\includegraphics[width=6.5in]{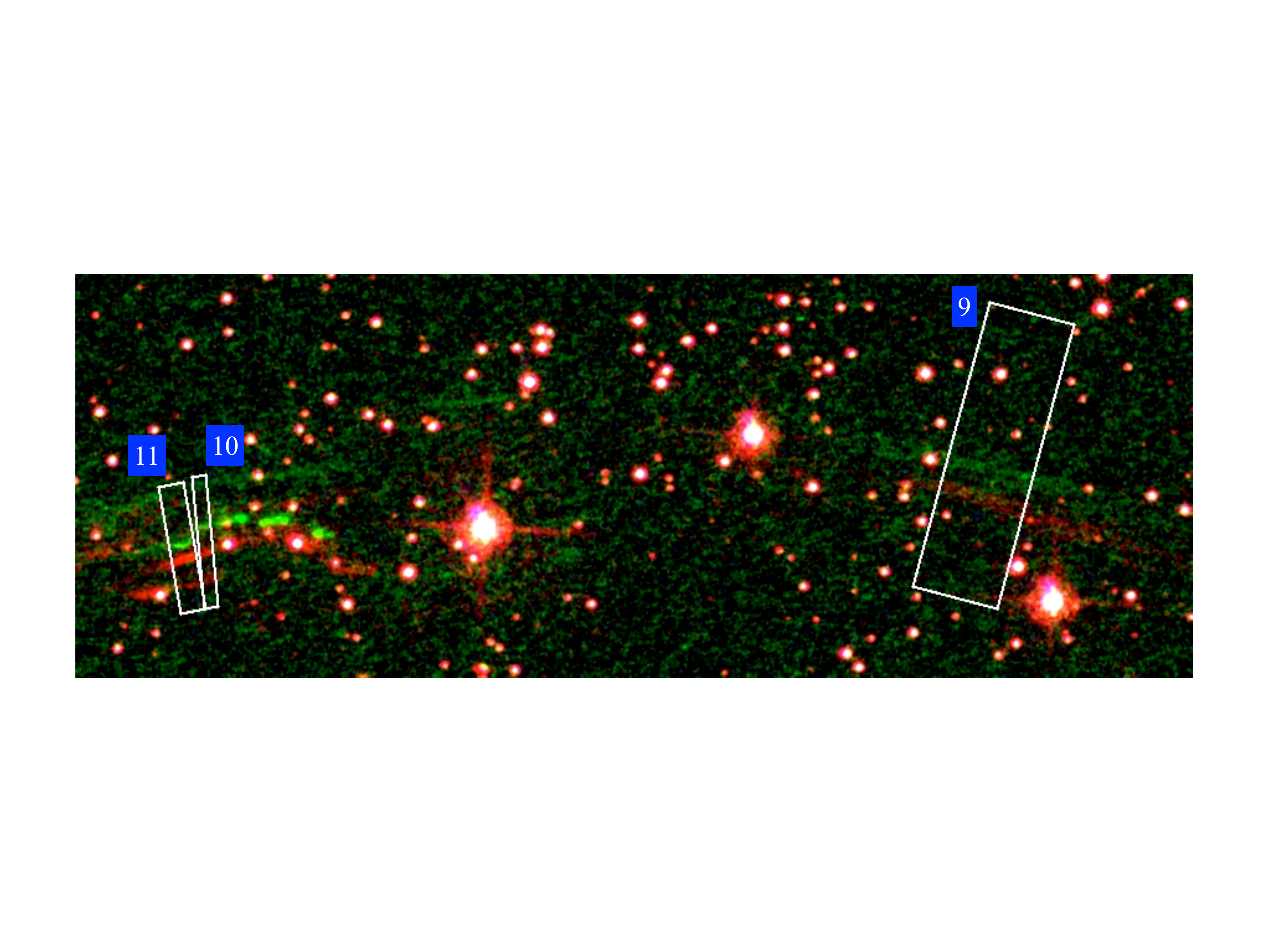}
\caption{The same as Figure \protect\ref{fbox1} for Box 3.  In this
case, the images have been smoothed by $2\times2$ pixel FWHM gaussian.
The image dimensions are about $36\arcsec\times 13\arcsec$.
\label{fbox3}}
\end{center}
\end{figure}

\begin{figure}
\begin{center}
\includegraphics[width=6.5in]{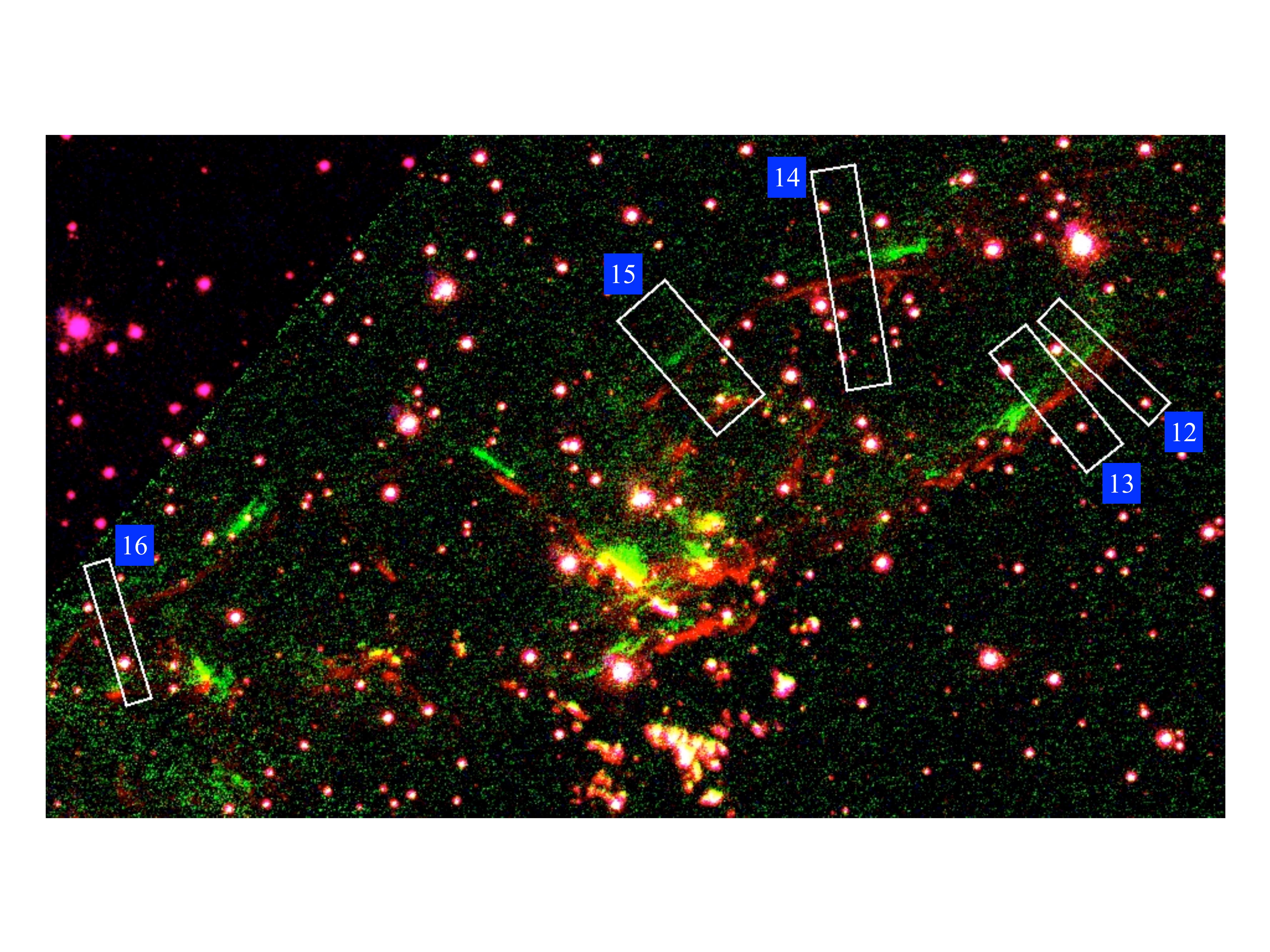}
\caption{The same as Figure \protect\ref{fbox1} for Box 4.  The
image dimensions are about $38\arcsec\times 22\arcsec$.
\label{fbox4} }
\end{center}
\end{figure}

Figs.~\ref{fbox1}--\ref{fwedgefil} show three-color images of the
regions ``Box 1'', ``Box 2'', ``Box 3'', ``Box 4'', ``Ejecta Knot'',
``Central Knots'' and ``Wedge Filaments''.  In each of these figures,
the epoch 1 ACS F658N (H$\alpha$+\nii\,) image is shown in red, the
epoch 1 ACS F660N (\nii\ only) image in blue and the epoch 2 WFC3
F656N (H$\alpha$ only) in green.  In the epoch 1 observations, the
lack of emission in the \nii--only image allows us to distinguish
those filaments emitting only H$\alpha$.  Any \nii\ emission fainter
than the F660N image background limit that may be included in the
F658N data will have no effect on the measured displacements.  In
the figures, the linear structures in red are the positions of the
Balmer filaments in the epoch 1 images, and those in green are their
positions in the epoch 2 data.  Radiative shocks appear white or
yellow, and stars appear white.  Fig.~\ref{fbox1} shows the region
with the brightest radiative shocks.  The edges to the knots and
clumps appear red due to the much higher signal to noise of the ACS
data compared with the WFC3 data, and due to the scaling which has
been chosen to show the faint features.

The locations for measuring proper motions were chosen by eye, based
on epoch 1, epoch 2 and combined 3-color images.  The primary
criterion for the selection was that linear features are identifiable
at both epochs and are translated more or less in a perpendicular
direction between epochs.  For the final list of positions measured,
we required a section of the filament at least 10 pixels wide free
of contamination by stars or radiative knots in both epochs.  The
direction of the cross-cuts was also chosen by eye, but using an
IDL routine where perpendicular lines were displayed and could be
aligned with the filaments.  The cross-cuts used are displayed in
Figs.~\ref{fbox1}--\ref{fwedgefil}.  For the primary shock front
(Boxes 1, 2, 3, 4) these are labeled numerically starting with the
westernmost position, and moving counter-clockwise around the remnant
periphery.

The features corresponding to filaments are clearly identifiable
in the cross-cuts.  Two examples are shown in Fig.~\ref{fplots} -
the top panel shows filament position 7 (Box 2), and the bottom
panel shows the very faint filament 9 (Box 3).  The alignment of
the stars are also evident in these plots.  The profiles from the
two epochs were cross-correlated for a range of lags within some
reasonable range.  The cross-correlation vs.\ lag curve was fit by
a quadratic, and the value of the lag corresponding to the
cross-correlation minimum was taken to be the filament displacement.
(The implementation was carried out using standard routines in IDL.)
In order to check the error due to placement, we varied the angle
of the cross-cuts by small amounts, and recalculated the lag
minimizing the cross-correlation function and found that the
measurements are accurate to about 0.5 pixels.

The proper motions in units of \arcsec/yr were determined from the
measured displacement, using 0.0396 \arcsec/pixel and an elapsed
time of 9.85~yr between the two epochs.  The uncertainty in the
proper motion for individual measurements is about 0.002 \arcsec/yr.
The proper motions, along with other filament properties, are
presented in Table~\ref{tblpm}.  The first column is the filament
label and the second gives the co-ordinates of the filament position
in the epoch 1 ACS F658N image.  The third column gives the box
widths (in pixels) along the filaments used in generating the
cross-cuts.  The fourth column contains the measured proper motions
in \arcsec/yr.  The fifth and sixth columns contain the peak
intensities and widths of the H$\alpha$ filaments in the epoch 1
data.  These values were obtained by fitting gaussians plus a linear
background to the extracted profiles.  The peak intensities in the
table are in units of $10^{-16}$ \sbu.  The Data Numbers
($\bar{e}$/s/pixel) were converted to physical units by multiplying
them by the inverse sensitivity, $1.96\times 10^{-18}$
erg/cm$^{2}$/\AA/$\bar{e}$ and the filter width, 78.0\AA\ (ACS
Instrument Handbook, Table 5.1), and dividing by the area of a
pixel, $1.57\times 10^{-3}$\,arcsec$^{2}$.  An important caveat is
that the values for the peak intensities are only approximate since
the conversion assumes a constant flux across the filter and does
not account for the shapes and radial velocities of the H$\alpha$
line.  Most of the non-radiative shocks have widths (FWHM) of order
35\AA, but a few are as wide as 80\AA, and since about half the
flux is in the broad component of the line, in these cases the
filter profile may reduce the flux by up to 20\%.


\section{Discussion}

\subsection{Distance to Kepler's SNR}

One of the goals of measuring Balmer filament proper motions is to
combine them with independently determined shock velocities, for
instance from H$\alpha$ line widths \citep{chevalier78, ghavamian07,
vanadelsberg08}, and calculate the distance to Kepler.  Under the
assumption that we are viewing a shock edge-on, and that the observed
filament traces the shock front at both epochs, the distance is
given by
\begin{equation}
D(pc)=\frac{0.21\,v_{shock}(km\,s^{-1})}{\Delta\Theta(\arcsec)/ \Delta\,t(yr)}
\end{equation}

\noindent and in the remainder of this paper, we will assume that this
relationship is valid.

The H$\alpha$ line widths of Balmer filaments have been measured
in Kepler by \citet{fesen89} and \citet{blair91}.  Both studies
targeted two positions, one in the NE, near filaments 12 and 13,
and the other in the NW, near filaments 7 and 8.  \citet{fesen89}
obtained spectra with the slits running east to west, thereby cutting
across the filaments.  They reported FWHM values of $1750\pm\,
200$\,\kms\ for the broad H$\alpha$ component at both locations.
\citet{blair91} positioned the slits along the filaments, and
estimated H$\alpha$ broad component line widths of about 1620\,\kms\
for the NE filament and 2330\,\kms\ for the NW filament.  The
significantly larger line width measured for the NW position by
\citet{blair91} may be due to contamination of the line profile by
emission from other filaments in this complex region.  Van Adelsberg
et al.\ (2008) have used the H$\alpha$ broad component width reported
by \citet{fesen89} to infer a shock velocity of $1589^{+191}_{-182}$\,\kms\
(see their Table 1).  \citet{morlino13} give a different relation
between shock speed and H$\alpha$ width, but the difference is only
significant for higher shock speeds.

For a given distance to the remnant, the shock velocity is proportional
to the proper motion.  The proper motions at positions 12 and 13
are the same -- 0.06\arcsec, and those at positions 7 and 8 are
about 10\% and 25\% larger (Table~\ref{tblpm}).  Thus, the shock
velocities among the positions will have a corresponding dispersion,
which in turn should be reflected in the broad component widths of
the H$\alpha$ emission.  Unfortunately, due to the relatively large
error-bars in the spectra, the differences between the NE and NW
filaments were not measurable, and in fact were reported to be the
same value \citep{fesen89}.  Therefore we are forced to make some
assumption about which value of proper-motion corresponds best to
the reported shock velocity.  We assume that the shock velocity,
1589\,\kms\ applies to the average proper motion measured at the
four positions, 7, 8, 12 and 13, which equals 0.06575\,\arcsec\,/yr.
Using these values we find D=5.1$^{+0.8}_{-0.7}$~kpc, where the
error intervals are based on the uncertainties quoted above for the
shock velocity and the error on a single measurement of the proper
motion, which is 0.002\,\arcsec\,/yr (see \S4).

Our newly derived distance to Kepler is significantly higher than
our earlier estimate \citep{sankrit05}, which was based on the
proper motion measurement of the NW filament (near positions 7 and
8 of this study).  The difference is primarily due to the higher
proper motion, $0.09\pm0.02$\arcsec\,/yr, used in the earlier study
compared to the value used here.  The current measurement of the
NW filament proper motion is $\approx0.07$\arcsec\,/yr, which is
consistent (within the errors) with the earlier value.  It is
unlikely that the change in proper motion is due to deceleration
of the filament.  The decrease of the shock velocity by about 28\%
in 10 years would require an increase in the pre-shock density of
about 65\% in the same period.  The gradient would have to be
sufficiently smooth that the filament retains its linear morphology
and does not show any significant changes in brightness.  Typically,
density contrasts are likely to be more clumpy - as evidenced by
the example discussed in \S5.2.1.  The errors on the proper motion
measurements reported in this paper are an order of magnitude lower
than in the earlier study.  Furthermore, the shock velocity adopted
here, about 4\% lower than used in our earlier study, benefits from
the recent theoretical calculations of \citet{vanadelsberg08}.  We
conclude that our new value for the distance is more reliable than
that obtained by \citet{sankrit05}.

\subsubsection{Caveat: Oblique Shocks}

Recently, \citet{shimoda15} pointed out that shocks in a clumpy
medium are rippled and mostly oblique, and that the obliquity can
affect the relationship between the shock proper motion and the
H$\alpha$ line width.  Their 3D hydrodynamic simulations indicated
that the downstream temperatures as indicated by the H$\alpha$ line
widths are on average 24\% below those expected from the proper
motions, which is to say that the H$\alpha$ line width is 11\% lower
than would be expected from the proper motion.  Hydrodynamic
simulations of the effect of shock rippling on the H$\alpha$ profiles
could in principle be misleading in that shocks are typically spread
over 3 to 5 cells, and the shock position must be chosen far enough
back that the final temperature has been reached.  However,
\citet{shimoda15} have been careful to avoid such problems.

The prediction that the shock speeds corresponding to the proper
motion exceed those derived from the H$\alpha$ line width is somewhat
counter-intuitive, because energy conservation implies that the
shock energy is converted to the sum of thermal energy and turbulent
kinetic energy, and if both thermal and turbulent velocities are
isotropic the relation between line width and shock speed must be
the same as in a planar shock.  \citet{shimoda15} attribute the
lower line widths to anisotropic turbulence.  Another possibility
is that the shock moves most rapidly through the low density regions
and ``heals'' after it crosses a small high density region.  In
that case, the proper motion would correspond to the faster shock
speed, while the H$\alpha$ line width would be an average of high
and low speeds.

It is somewhat surprising that the density contrast of a few percent
expected in the ISM at the relevant scales could produce as large
an effect as reported by \citet{shimoda15}, but the irregularity
of the shock surface and the presence of both radiative and
non-radiative shocks indicate a very high level of density fluctuations
in the CSM of Kepler. Thus the effects of shock rippling should be
much stronger than in remnants such as Tycho or SN1006.  One estimate
of the level of irregularity of the shock surface is the ratio of
the thickness of a filament to its length.  As seen in Figure 2,
the straight segments of the filaments are typically 5\arcsec\ long,
while the thicknesses listed in Table 1 are about 0.2\arcsec,
for typical ratios of 25.  By comparison, the H$\alpha$ filaments
in the NW section of SN1006 have lengths of arcminutes and thicknesses
of about 0.3\arcsec, for a ratio of over 200 \citep{raymond07}.
Moreover, the filament thickness in SN1006 corresponds to the angular
ionization length scale of H~I \citep{heng07}, while the angular ionization
length scale in Kepler is about 10 times smaller due to its higher
densities, lower shock speeds and larger distance.  Therefore, the
thickness of the SN1006 filaments is only an upper limit to the
amplitude of shock ripples, while in Kepler it may be due to either
curvature on the scale of the filament length or to smaller scale
rippling.

We conclude that because of the strong density fluctuations in thefuther
pre-shock medium of Kepler's SNR, care will be needed in using the
proper motions and shock speeds from H$\alpha$ profiles to obtain
the distance.  Some handle on the effects of rippling can be obtained
simply by looking at the scatter among distances derived from
different filaments.  We defer further analysis until we have a
larger set of H$\alpha$ line widths for the filaments whose proper
motion we have measured.

The distance of 5.1~kpc to Kepler will be a modest overestimate
if the shock velocity from the H$\alpha$ line width is lower than
that from proper motion.

\subsection{Shock Velocities and Intensities}

For a distance of 5.1~kpc the proper motion shock velocity is given
by: $v_{s}(km\, s^{-1})\approx 24300\times
\Delta\Theta/\Delta\,t$(\arcsec/yr).  The error of 0.5 pix in the
measured displacement of the filaments translates to an error in
the shock velocity of about 50\,\kms.  Fig.~\ref{fpkvs} shows a
plot of the H$\alpha$ fluxes per arcsecond length along the filament
versus shock speed.  The fluxes are derived from the Gaussian fits
(Table~\ref{tblpm}) as $I_{peak}\times FWHM\times 1.064$.

For a non-radiative shock with depth $L$ along the line of sight,
shock velocity $v_s$, pre-shock density $n_0$, and neutral fraction
$f_{neut}$, the flux (uncorrected for interstellar extinction) is
\begin{equation}
F = \frac{N_{H\alpha}\, n_0\, f_{neut}\, v_s\, L\, h \nu}{4 \pi}\
erg~cm^{-1}~s^{-1}~sr^{-1}
\end{equation}

\noindent where $N_{\rm{H}\alpha}$=0.25 is the number of H$\alpha$
photons per neutral H atom \citep{laming96} and $h\nu\,=3.03\times
10^{-12}$\,erg is the photon energy.  We use the reddening to Kepler,
E(B-V)=0.9 \citep{blair91} and R$_{V}$=3.1 to obtain an extinction
$A_{\rm{H}\alpha}=2.27$.  Using appropriate values for the parameters
and correcting for the interstellar extinction yields
\begin{equation}
F = 0.9\times 10^{-16} n_0 f_{neut} v_{1000} L_{17.6}\
erg~cm^{-2}~s^{-1}~arcsec^{-1}
\end{equation}

\noindent where $v_{1000}$ is the shock speed in units of 1000~\kms,
and $L_{17.6}$ is the line of sight depth in units of $4\times
10^{17}$~cm, corresponding to the 5\arcsec\ length typical of the
filaments.  The neutral fraction is not known, but has been estimated
at 0.1 for non-radiative shocks in SN1006 \citep{ghavamian02}, 0.8
for Tycho \citep{ghavamian01} and 0.06--0.2 for the Cygnus Loop
\citep{raymond13}.  In Kepler it is likely to be small in filaments
close to radiative shocks, but may vary significantly from one
location to another.

\subsubsection{The Main Shock}

The proper motion shock velocities for the main shock positions,
shown as diamonds in Fig.~\ref{fpkvs}, range from 1190~\kms\ (position
11) to 2280~\kms\ (position 9).  The fluxes of all but two of the
positions (2 and 5) lie between 0.44 and 3.24$\times\,10^{-16}$\,\flu.
Excluding these outliers, the median shock speed is 1690~\kms\ and
the median flux is 1.21$\times\,10^{-16}$\,\flu, which imply that
if $f_{neut}=0.1$ and $L_{17.6}=1$, then the shocks are encountering
densities of about 8\,\cc.  This is consistent with the post-shock
density of 42\,\cc\ derived from IR spectroscopy by \citet{williams12}
for a region in the northern part of the remnant (within Box 3 of
this paper).

Positions 2 and 5 have shock velocities of 1340 and 1410~\kms,
respectively, which are about 20\% lower than the median value, but
fluxes that are about 5.5 and 4.5 times the typical value.  There
is evidence for hotter dust in these regions from ISOCAM 14--16\,$\mu$m
data \citep{douvion01} and \spitzer\ IRS spectra \citep{williams12}.
The higher dust temperatures indicate higher densities.  Since the
shock velocities at positions 2 and 5 are typical, the brightness
and higher densities suggest that the driving pressures are
significantly higher than average at these locations.  Position 2
is close to the apex of a curved structure (Fig.~\ref{fbox1}) where
the blast wave seems to be wrapping around a cloud.  In this case,
significant pressure variations are expected \citep{borkowski97}.
There is no similar morphological evidence at position 5, but
significant pressure variations are expected even on small spatial
scales.  We may turn the argument around and argue that the shock
velocity and brightness of the filament at position 5 implies a
higher pressure at that location.

The variation in shock properties between positions 3 and 4 is
noteworthy.  They lie on the same filament and are separated by
about 0.7\arcsec\ (Fig.~\ref{fbox1}).  The shock speed at position
3 is 1770~\kms, and at position 4 it is 1970~\kms.  It is likely
that the shock is slower at position 3 due to a higher density clump
in the pre-shock gas.  If the driving pressure, $n_{0}v_{s}^{2}$
is the same at both locations, then the density contrast is about
25\%.  However, the flux at position 3 is almost 6 times higher
than at position 4.  A straightforward explanation is that the
neutral fraction in the dense clump is higher by a factor of 5
compared with its surroundings.

An extreme variation in shock velocity over a small scale occurs
between positions 10 and 11, which are separated by only 0.7\arcsec\
(Fig.~\ref{fbox3}).  The appearance of the region clearly shows
complex changes between epoch 1 and 2.  The shock velocities at the
two positions are 2140 and 1190~\kms, respectively.  Assuming
$f_{neut}=0.1$ and $L_{17.6}$=1 for both positions, equation 3
yields densities of 6.0 and 16.3\,\cc\ for positions 10 and 11.
These values are consistent with the driving pressure being the
same (to within 15\%) at the two positions.

\subsubsection{The Ejecta Knot}

\begin{figure}
\begin{center}
\includegraphics[width=6.5in]{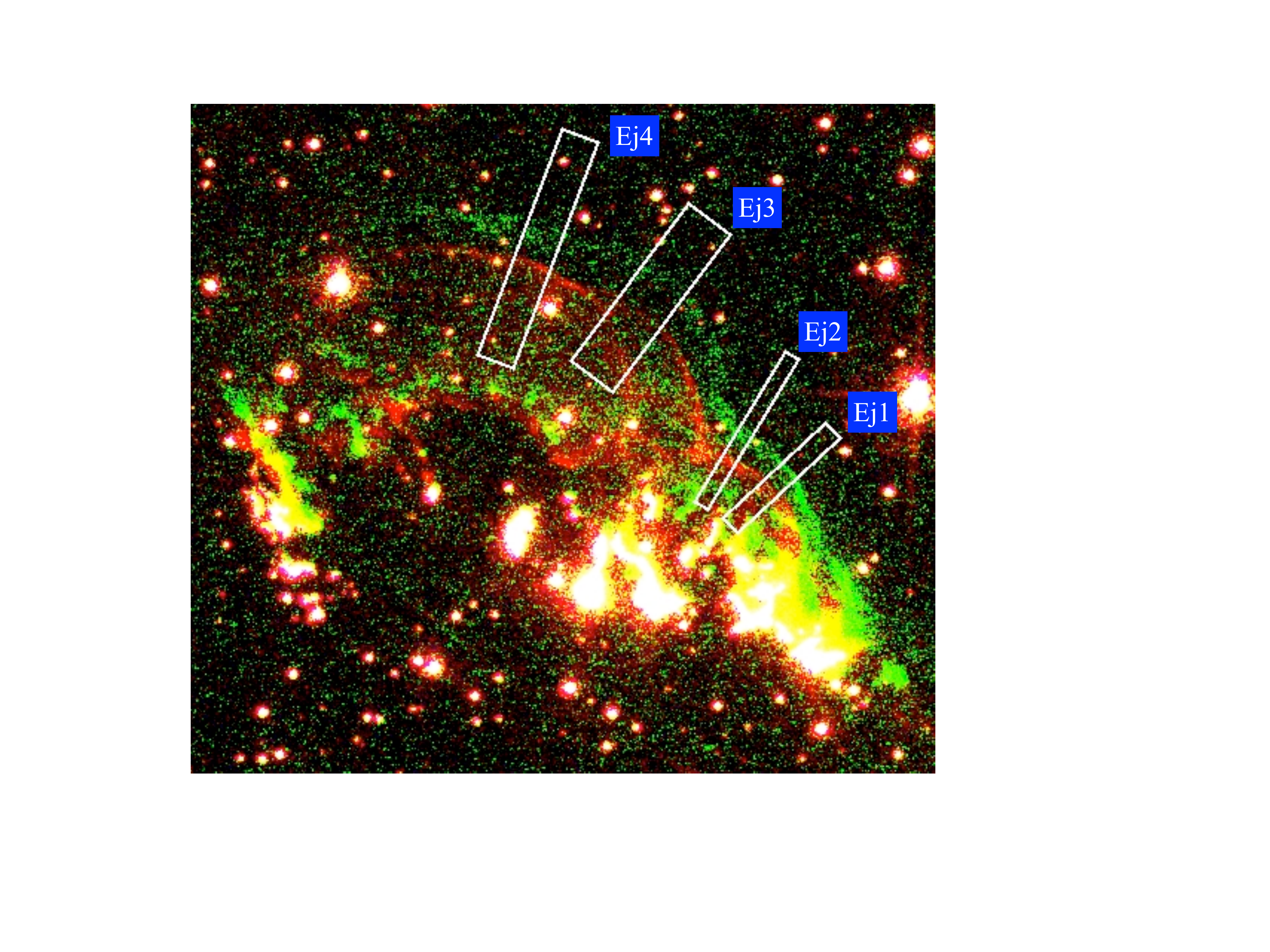}
\caption{The same as Figure \protect\ref{fbox1} for the Ejecta Knot.
The image dimensions are about $20\arcsec\times 18\arcsec$.
\label{fejknot} }
\end{center}
\end{figure}

The Ejecta Knot (Fig.~\ref{fejknot}) first appeared \textit{ca.}
1970 \citep{vandenbergh77} and brightened rapidly until by 1985 its
H$\alpha$+\nii\ intensity was comparable to the brightest regions
in the remnant \citep{dodorico86}.  The detailed structure of the
region was revealed by our first epoch HST images \citep{sankrit08}
as consisting of arc-shaped Balmer dominated filaments tracing a
shock front and trailing radiative knots due to Rayleigh-Taylor
like instabilities.  The clear correlation between the emitting
structures in the X-ray and in H$\alpha$ may be seen in Fig.~\ref{fejxha}.
The deep (750~ks exposure) \chandra\ data used for the image was
analyzed by \citet{reynolds07} and \citet{burkey13} who showed that
the emission was from a knot of ejecta rather than a dense clump
in the CSM.

Between epoch 1 and 2, the sharply defined filaments have become
more diffuse.  This is most clearly seen for Ej4 (Fig.~\ref{fejknot},
where the filament position is poorly defined in the 2013 image,
and consequently only an approximate value of the proper motion is
reported (Table~\ref{tblpm}).

Positions on the shock front ahead of the Ejecta Knot are shown as
thick crosses in Fig.~\ref{fpkvs}.  The fastest moving is position
Ej3, where the shock velocity is 2960~\kms, significantly higher
than any of the main shock positions.  The shock velocities at the
remaining three positions fall within the main shock range, and are
biased towards higher values.  The flux at Ej3 is
$1.03\times\,10^{-16}$\,\flu, and $L_{17.6}\approx 0.25$
(Fig.~\ref{fejknot}), which yield $n_{0}\,f_{neut}=1.5$\,\cc.  The
neutral fraction in the pre-shock gas ahead of the ejecta knot is
unknown, but could plausibly be higher than the value of 0.1 assumed
for the main shock.  The extreme case of $f_{neut}=1.0$ implies
that $n_{0}=1.5$\,\cc.

\subsubsection{The Central Knots and Wedge Filaments}

\begin{figure}
\begin{center}
\includegraphics[width=6.5in]{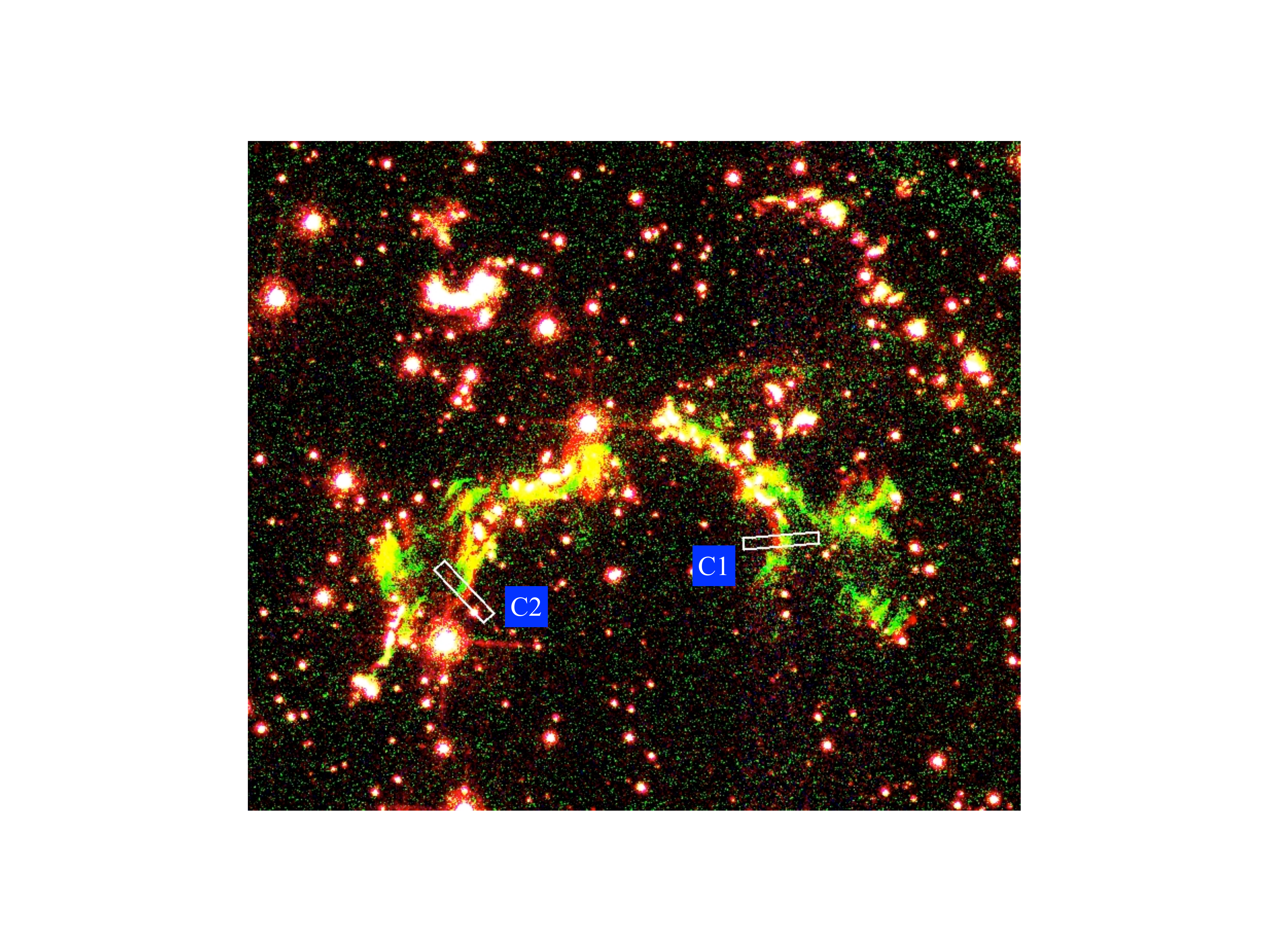}
\caption{The same as Figure \protect\ref{fbox1} for the Central
knots.  The image dimensions are about $30\arcsec\times 26\arcsec$.
\label{fcentknots} }
\end{center}
\end{figure}

\begin{figure}
\begin{center}
\includegraphics[width=6.5in]{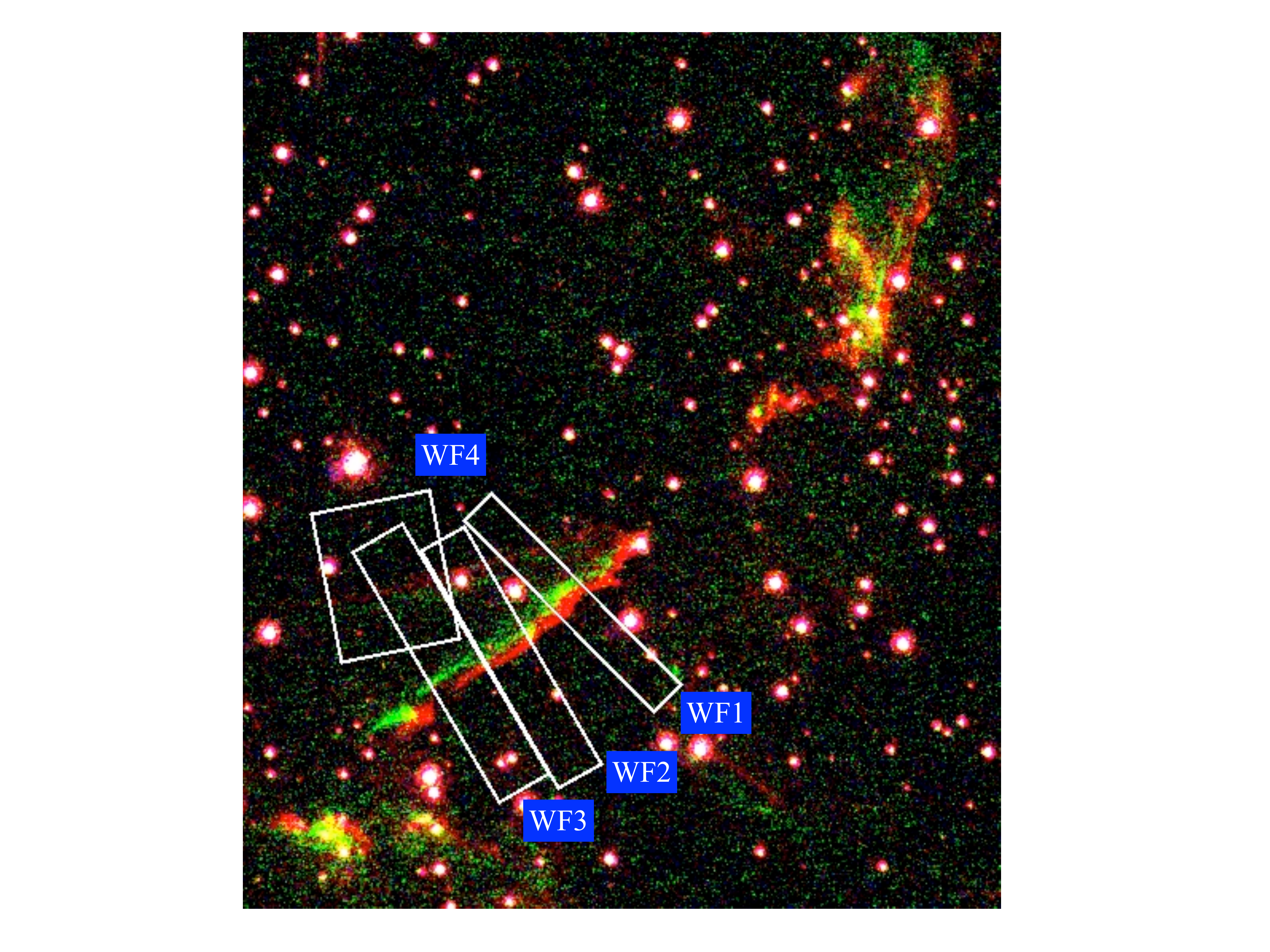}
\caption{The same as Figure \protect\ref{fbox1} for the Wedge
Filaments.  The image dimensions are about $19\arcsec\times 22\arcsec$.
\label{fwedgefil} }
\end{center}
\end{figure}

\begin{figure}
\begin{center}
\includegraphics[height=6.5in]{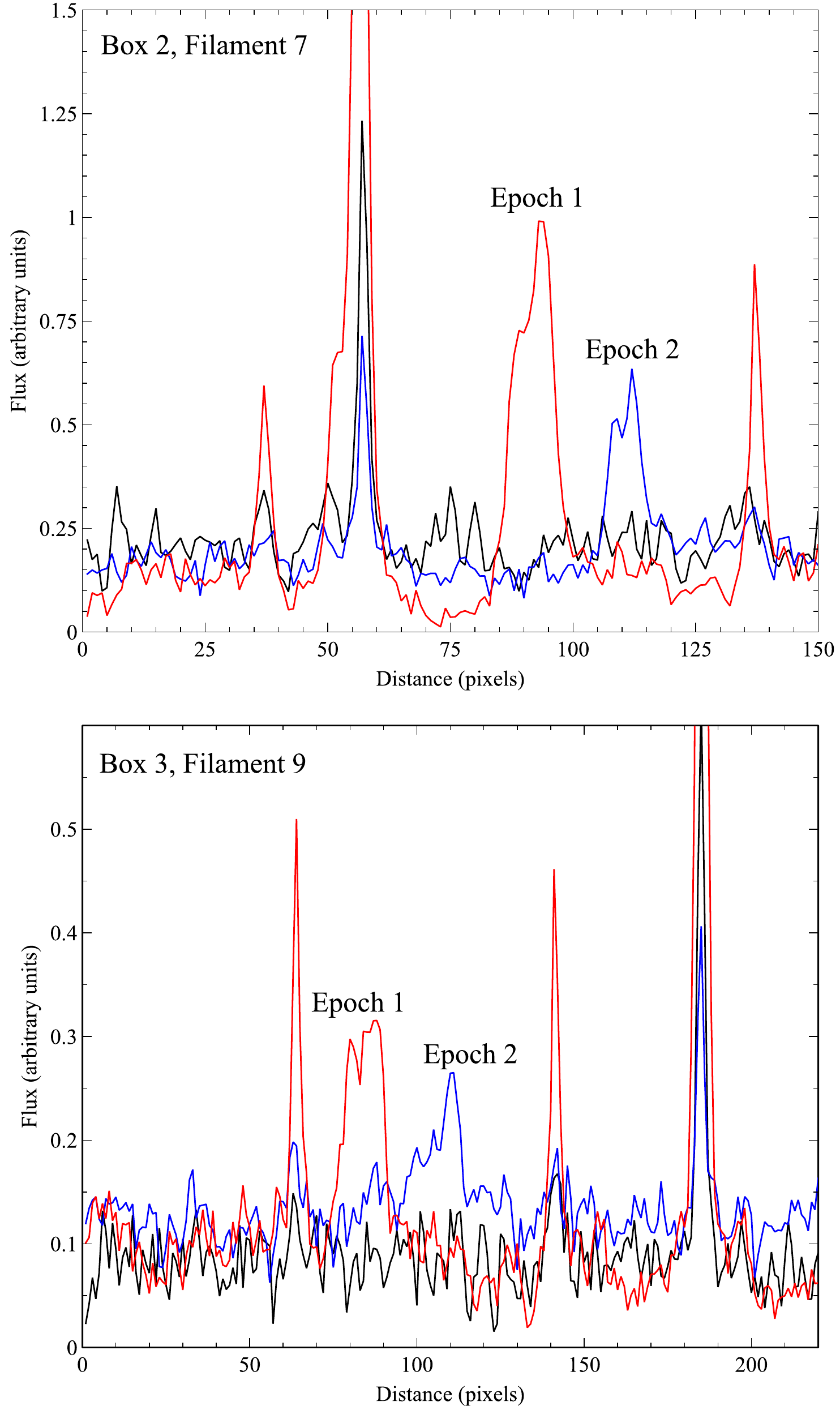}
\caption{H$\alpha$ emission profiles across Balmer line filaments.
Top panel: filament 7 in Box 2 (Fig.~\protect\ref{fbox2}); bottom
panel: filament 9 in Box 3 (Fig.~\protect\ref{fbox3}).  In red is
the epoch 1 ACS F658N emission, in blue is the epoch 2 WFC3 F656N
emission, and in black is the epoch 1 ACS F660N emission, which is
essentially zero except at the location of the stars.  The labels,
``Epoch 1'' and ``Epoch 2'' in each of the panels show the locations
of the Balmer filament in 2003 and 2013, respectively.
\label{fplots} }
\end{center}
\end{figure}

\begin{figure}
\begin{center}
\includegraphics[width=6.0in]{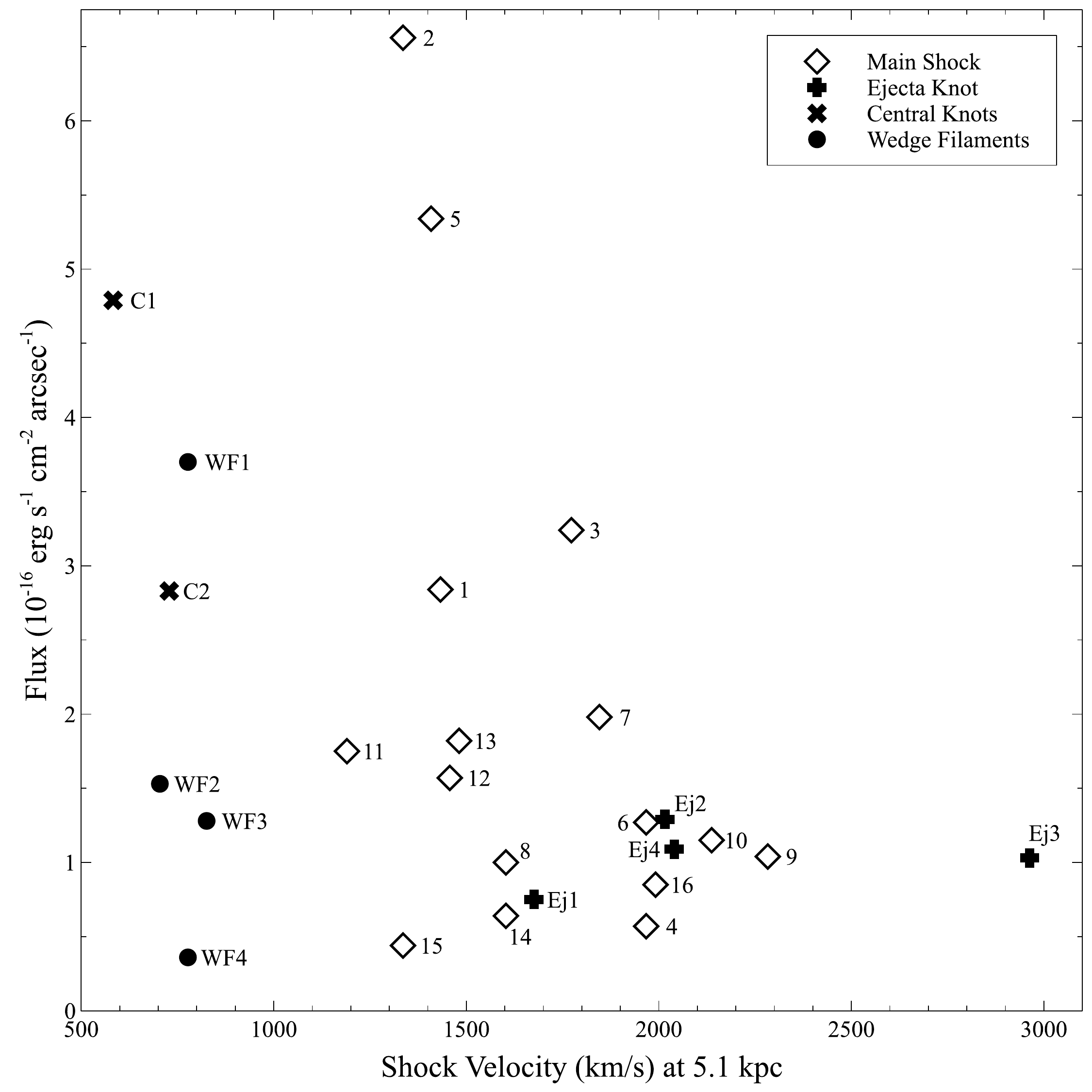}
\caption{The H$\alpha$ emission peak intensities are plotted against
the proper motion shock velocities obtained assuming a distance of
5.1~kpc to Kepler.  The intensity and proper motion values are from
Table 1, as are the labels used for each point.
\label{fpkvs} }
\end{center}
\end{figure}

\begin{figure}
\begin{center}
\includegraphics[width=7.5in]{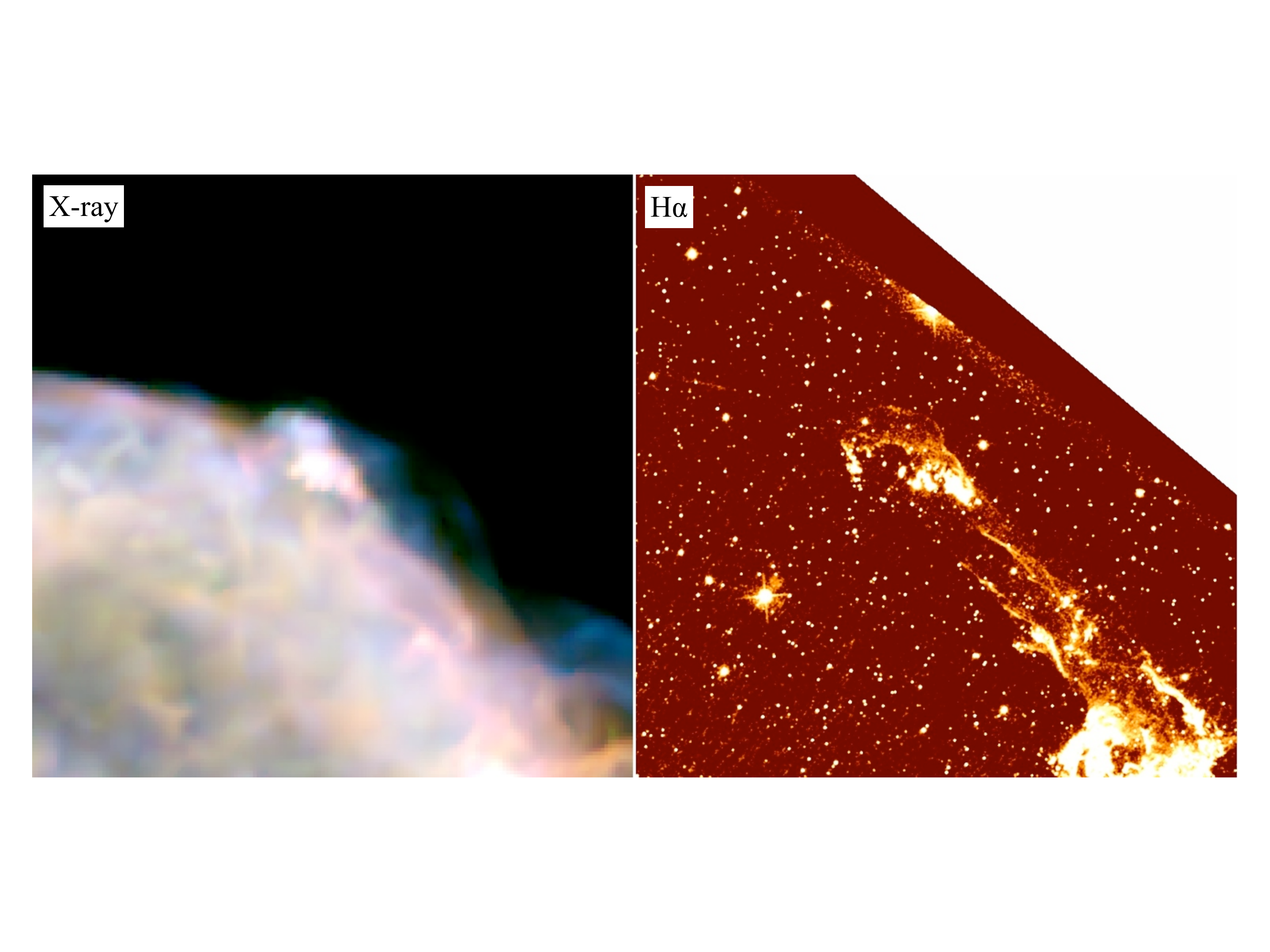}
\caption{The region around the ejecta knot.  Left: A smoothed
three-color \chandra\ image with 0.3--0.72~keV emission in red,
0.72--1.7~keV in green and 1.7--7.0~keV in blue. Right: HST/WFC3
F656N (H$\alpha$) image.  The field of view shown is an
82\arcsec$\times$\,88\arcsec.
\label{fejxha} }
\end{center}
\end{figure}

\begin{figure}
\begin{center}
\includegraphics[width=7.5in]{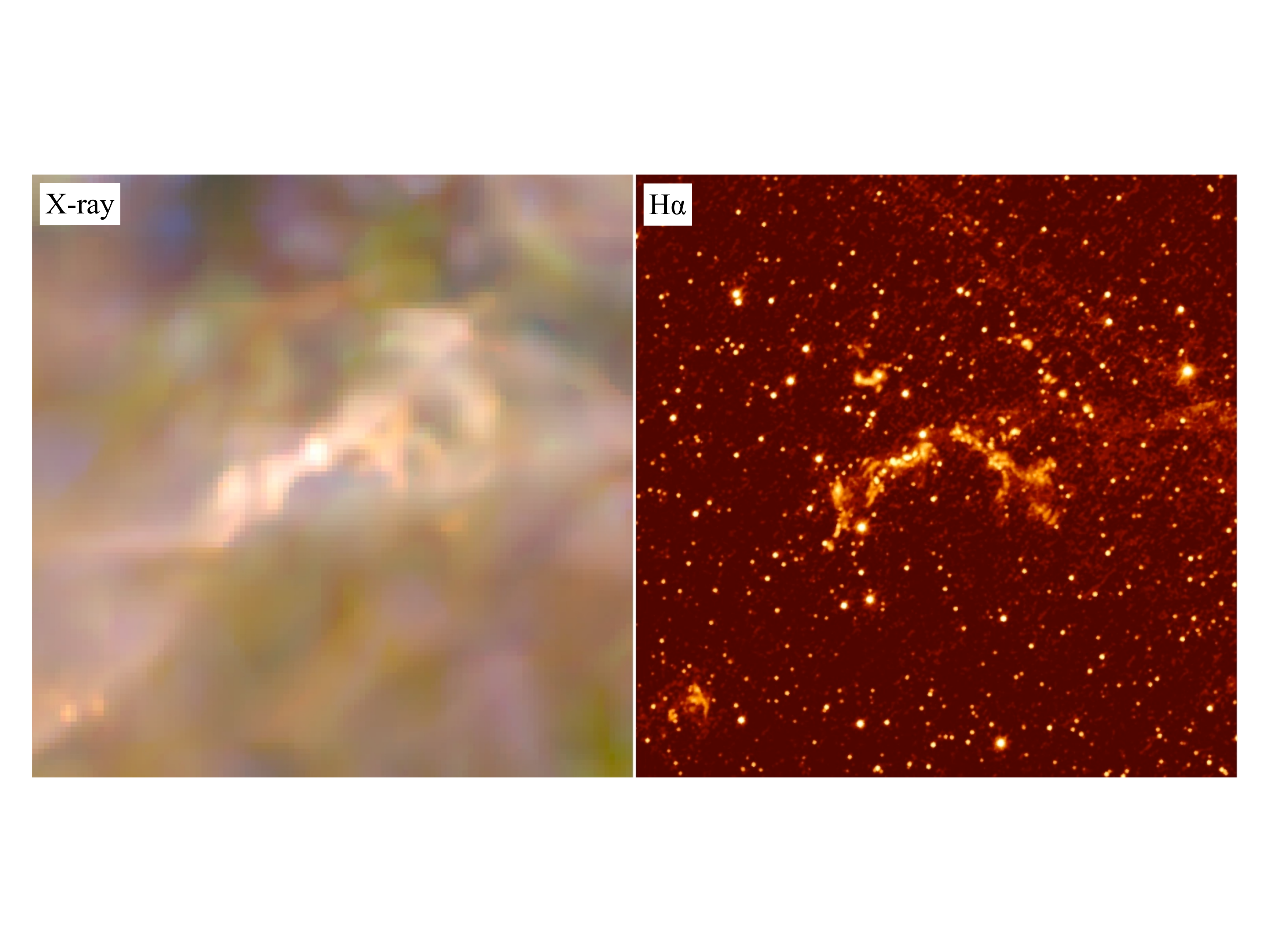}
\caption{The same as Figure \protect\ref{fejxha}, for a $50\arcsec\times
50\arcsec$\ region around the Central Knots.
\label{fcentxha} }
\end{center}
\end{figure}

The central knots comprise Balmer dominated filaments due to
non-radiative shocks and radiatively shocked knots interspersed
among each other within a small area (Fig.~\ref{fcentknots}).  The
optical emission is blue-shifted relative to the mean radial velocity
of Kepler \citep{blair91}.  The X-ray emission from the region
correlates well with the H$\alpha$ emission (Fig.~\ref{fcentxha}).
\citet{burkey13} have shown that the X-ray emission from the region
is primarily from circumstellar material.  This may be seen by
comparing the X-ray image of the Central Knots region with that of
the NE limb shown in Fig.~\ref{fejxha}.  Both images have been
scaled identically, and the redder appearance of the central knots
is due primarily to oxygen lines, characteristic of the CSM.

The H$\alpha$ emission morphology is complex, and although a number
of changes between the two epochs are evident (Fig.~\ref{fcentknots}),
unambiguously identifying filaments that trace the motion of shock
fronts is difficult.  We have isolated two positions, C1 and C2,
that plausibly do represent real filament motion.  We note that the
filaments lying to the southwest of C1 that appear green in
Fig.~\ref{fcentknots} are not newly emerged in epoch 2; they are
clearly visible in the epoch 1 data, but due to the stretch used
in the display are not seen in the image presented.

The wedge filaments (Fig.~\ref{fwedgefil}) are a pair of linear
Balmer filaments that define a wedge-shaped structure, which is
located well inside the projected boundary of the remnant
(Fig.~\ref{fxacsirfull}).  The morphology suggests that the primary
shock encountered an inward protrusion and wrapped around its sides,
and the filaments are the tangencies of the shock front along the
line of sight.  Proper motions could be measured at three adjacent
locations (WF1, WF2 and WF3) on the brighter southern filament,
whereas a very wide box was necessary to get sufficient counts for
the measurement on the fainter northern side.

In Fig.~\ref{fpkvs} thick `X's show the positions in the central
region, and filled circles those on the wedge filaments.  The shock
velocities at positions C1 and C2 are 580 and 730~\kms, respectively,
and the fluxes are 4.79 and 2.73$\times 10^{-16}$\,\flu.  Assuming
$f_{neut}$=0.1, and $L_{17.6}=0.2$ based on the 0.8\arcsec\ filament
length (Fig.~\ref{fcentknots}), we obtain pre-shock densities of
about 460 and 220\,\cc\ at the two locations.

The shock velocities for the wedge filament positions all lie in a
narrow range 700--930~\kms.  The fluxes along the southern side
show a systematic decrease away from the thin edge of the wedge,
falling from 3.70 to 1.28$\times 10^{-16}$\,\flu\ between WF1 and
WF3.  The northern side is fainter still with a flux 0.36$\times
10^{-16}$\,\flu.  The gradient along the southern side may be due
to a density gradient.  The difference between southern and northern
sides is probably caused by the geometry of the wedge.  One scenario
that explains the contrast in brightness is that the northern shock
front is inclined to the line of sight, and therefore the length
through the emitting region is small.  This would also imply that
the proper motion shock speed is an underestimate of the true shock
speed.  The current data are not sufficient to distinguish between
this and other possibilities.

\citet{burkey13} have shown that the regions around the central
knots and wedge filaments are dominated by material from the
progenitor CSM (see their Fig.~2).  Their model of SN ejecta with
an exponential profile expanding into an azimuthally varying stellar
wind results in a torus of shocked CSM occupying the equatorial
plane.  In the context of this model, the Balmer filaments in the
central knots region and the wedge filaments are shocks being driven
into this torus.  The shock velocity and H$\alpha$ intensity
measurements thus provide quantitative information that can constrain
more detailed models of the shock-CSM interaction in Kepler's SNR.

\section{Concluding Remarks}

We have presented HST images of Kepler's SNR obtained approximately
ten years apart and measured the proper motion of the Balmer
filaments, which trace non-radiative shocks.  This study represents
a significant improvement over that of \citet{sankrit05} -- in the
current study 26 filament positions were accessible for measurement
as compared with two in the earlier work, and the statistical errors
on the measurements were reduced by an order of magnitude.  We used
the average proper motion for four filament positions and the shock
velocity reported in the literature to obtain a distance of 5.1~kpc
to Kepler, which is about 30\% higher than the value obtained by
\citet{sankrit05}.  The higher value reduces, but does not altogether
eliminate, the tension between the distance obtained using Balmer
filament proper motions and that required by the SN/CSM models of
\citet{chiotellis12} and \citet{patnaude12}, who required a 
distance greater than about 6~kpc.

Using our derived value of the distance, we have calculated the
proper motion shock velocities of the filaments.  We  have estimated
the densities in the pre-shock gas from these velocities and the
H$\alpha$ fluxes of the filaments.  The distribution of shock
velocities and pre-shock densities have been discussed for three
components: (i) the main shock along the remnant perimeter, (ii)
the shock driven ahead of the ejecta knot, and (iii) the shocks
near the center and in the projected interior of the remnant.

Our revised distance to Kepler, and the shock velocity and density
estimates provide information that can strongly guide and constrain
hydrodynamic models of the supernova blast wave interaction with
the CSM around the progenitor star.

\acknowledgments

We thank Derek Hammer (previously of STScI) for his expertise with
the AstroDrizzle package and providing us with the final aligned,
CTE-corrected images used for analysis.  Figures 9 and 10 were made
using Veusz, a scientific plotting package written by Jeremy Sanders.
This work was supported in part by NASA Grant HST-GO-12885 awarded
to the Universities Space Research Association and to the Johns
Hopkins University.

Facilities: 
\facility{Hubble Space Telescope.}


\clearpage

\newpage

\newpage

\newpage

\newpage

\newpage

\newpage

\newpage

\newpage

\newpage

\newpage

\newpage

\clearpage

\begin{deluxetable}{cccccc}

\tablecaption{Balmer Filament Properties \label{tblpm} }
\tablewidth{0pt}
\tablehead{
\colhead{Filament} 
  & \colhead{$\alpha_{J2000}$ ; $\delta_{J2000}$}
  & \colhead{Box Width} 
  & \colhead{$\Delta\Theta/\Delta\,t$}
  & \colhead{I$_{peak}$} 
  & \colhead{FWHM}
}
\startdata

   1 & 17:30:35.37 ; $-$21:29:01.1 & 10.5 & 0.059 & 15.6 & 0\farcs17 \\
   2 & 17:30:35.73 ; $-$21:28:58.2 & 14.2 & 0.055 & 20.5 & 0\farcs30 \\
   3 & 17:30:36.14 ; $-$21:28:46.2 & 10.4 & 0.073 &  7.8 & 0\farcs39 \\
   4 & 17:30:36.15 ; $-$21:28:45.5 & 19.3 & 0.081 &  3.9 & 0\farcs14 \\
\\
   5 & 17:30:37.03 ; $-$21:28:23.4 & 19.0 & 0.058 & 15.6 & 0\farcs32 \\
   6 & 17:30:37.45 ; $-$21:28:09.7 & 30.2 & 0.081 &  5.9 & 0\farcs20 \\
   7 & 17:30:37.54 ; $-$21:28:08.5 & 29.8 & 0.076 &  8.8 & 0\farcs21 \\
   8 & 17:30:37.65 ; $-$21:28:07.5 & 44.2 & 0.066 &  4.9 & 0\farcs19 \\
\\
   9 & 17:30:40.25 ; $-$21:27:48.1 & 70.5 & 0.094 &  2.0 & 0\farcs50 \\
  10 & 17:30:42.03 ; $-$21:27:49.9 & 10.9 & 0.088 &  4.9 & 0\farcs22 \\
  11 & 17:30:42.08 ; $-$21:27:50.1 & 20.5 & 0.049 &  7.8 & 0\farcs21 \\
\\
  12 & 17:30:43.04 ; $-$21:27:55.8 & 24.2 & 0.060 &  2.9 & 0\farcs50 \\
  13 & 17:30:43.15 ; $-$21:27:57.0 & 37.1 & 0.061 &  5.9 & 0\farcs29 \\
  14 & 17:30:43.62 ; $-$21:27:53.1 & 35.8 & 0.066 &  2.9 & 0\farcs21 \\
  15 & 17:30:43.97 ; $-$21:27:55.8 & 50.7 & 0.055 &  2.0 & 0\farcs21 \\
  16 & 17:30:45.31 ; $-$21:28:03.8 & 20.4 & 0.082 &  2.0 & 0\farcs41 \\
\\
 Ej1 & 17:30:37.97 ; $-$21:27:57.4 & 14.1 & 0.069 &  3.9 & 0\farcs18 \\
 Ej2 & 17:30:38.05 ; $-$21:27:56.7 & 10.6 & 0.083 &  6.8 & 0\farcs18 \\
 Ej3 & 17:30:38.21 ; $-$21:27:52.7 & 34.9 & 0.122 &  2.0 & 0\farcs50 \\
 Ej4 & 17:30:38.42 ; $-$21:27:51.4 & 25.2 & $\sim$0.08\tablenotemark{a} & 5.9 & 0\farcs18 \\
\\
 C1  & 17:30:39.70 ; $-$21:29:23.4 & 11.0 & 0.024 & 11.7 & 0\farcs39 \\
 C2  & 17:30:40.56 ; $-$21:29:25.5 & 13.1 & 0.030 &  9.8 & 0\farcs27 \\
\\
 WF1 & 17:30:43.93 ; $-$21:29:44.9 & 23.7 & 0.032 & 15.6 & 0\farcs22 \\
 WF2 & 17:30:44.04 ; $-$21:29:46.2 & 30.6 & 0.029 &  8.8 & 0\farcs16 \\
 WF3 & 17:30:44.12 ; $-$21:29:46.8 & 35.0 & 0.034 &  4.9 & 0\farcs25 \\
 WF4 & 17:30:44.25 ; $-$21:29:44.6 & 75.8 & 0.032 &  2.0 & 0\farcs18 
\enddata

\tablecomments{The columns are as follows -- (1) filament label
used in the text and figures, (2) approximate co-ordinates of the
filament in the 2003 ACS image, (3) width, in pixels, of the extraction
boxes, (4) filament proper motion, in arcsec/yr, (5) Peak intensity
of the filament profile in the Epoch 1 data in units of $10^{-16}$
\sbu\ and (6) full-width-half-maximum of the filament profile in
the Epoch 1 data.}

\tablenotetext{a}{Approximate value; the uncertainty in the measured
displacement is $\sim$2 pixels.}

\end{deluxetable}


\end{document}